\documentclass[aps,prx,twocolumn,showpacs,superscriptaddress,groupedaddress,longbibliography]{revtex4-1}  

\usepackage{graphicx}  
\usepackage{dcolumn}   
\usepackage{bm}        
\usepackage{amssymb}   
\usepackage{color}  
\usepackage{amsmath}
\usepackage{hyperref}
\usepackage{xcolor}

\hypersetup{
    colorlinks,
    linkcolor={blue!50!black},
    citecolor={blue!50!black},
    urlcolor={blue!80!black}
}
\begin{document}

\widetext

\title{Thermoelectric Scanning Gate Interferometry on a Quantum Point Contact}

\author{B. Brun$^{1*}$, F. Martins$^{1*}$, S. Faniel$^1$, A. Cavanna$^4$, C. Ulysse$^4$, A.Ouerghi$^4$, U. Gennser$^4$, 
D. Mailly$^4$, P. Simon$^5$, S. Huant$^2$, M. Sanquer$^3$, H. Sellier$^2$, V. Bayot$^1$ \& B. Hackens$^1$}

\affiliation{
$^1$IMCN/NAPS, Universit\'e catholique de Louvain (UCLouvain), B-1348 Louvain-la-Neuve, Belgium \\
$^2$Universit\'e Grenoble Alpes, CNRS, Institut N\'eel, 38000 Grenoble, France\\
$^3$Universit\'e Grenoble Alpes, CEA, INAC-Pheliqs, 38000 Grenoble, France \\
$^4$Centre de Nanosciences et Nanotechnologies (C2N) CNRS, Route de Nozay, F-91460 Marcoussis, France \\
$^5$Laboratoire de Physique des Solides, B\^atiment 510, Universit\'e Paris Sud, F-91405 0rsay, France \\
$^*$Contributed equally to this work
}

\date{\today}

\begin{abstract}
We introduce a new scanning probe technique derived from scanning gate microscopy (SGM) in order to 
investigate thermoelectric transport in two-dimensional semiconductor devices. 
The thermoelectric scanning gate microscopy (TSGM) consists in measuring the thermoelectric voltage induced 
by a temperature difference across a device, while scanning a polarized tip that locally changes the potential landscape.
We apply this technique to perform interferometry of the thermoelectric transport in a quantum point contact (QPC).
We observe an interference pattern both in SGM and TSGM images, and evidence large differences between the two signals in the low density regime of the QPC. 
In particular, a large phase jump appears in the interference fringes recorded by TSGM, which is not visible in SGM.
We discuss this difference of sensitivity using a microscopic model of the experiment, based on the contribution from a resonant level inside or close to the QPC.
This work demonstrates that combining scanning gate microscopy with thermoelectric measurements  offers new information 
as compared to SGM, and provides a direct access to the derivative of the device transmission with respect to energy, both in amplitude and in phase.
\end{abstract}

\maketitle

\section{Introduction}
In the context of emerging quantum technology and in view of the increasing care for energy
harvesting, thermoelectric transport in nanomaterials and nanodevices has recently regained interest \cite{Dresselhaus-2007,Jiang-2016}.
This has led to novel quantum thermal devices such as caloritronics interferometers \cite{Giazotto-2012}. 
The ability to accurately measure heat transport in two-dimensional systems \cite{Jezouin-2013} and atomic junctions \cite{Cui-2017} improved our global understanding 
of quantum thermodynamics, and shed light on mechanisms at play in complex many-body problems \cite{Banerjee-2017, Sivre-2017, Dutta-2017}.
Investigating these thermal effects at the local scale is challenging, but lots of efforts are also made in this direction. 
As an example, heat dissipation was recently mapped inside a graphene nanodevice with unprecedented thermal and spatial resolutions \cite{Halbertal2016},
unveiling new mechanisms responsible for current-to-heat conversion in graphene \cite{Halbertall-2017}.

Here we introduce a new scanning probe technique based on the Seebeck effect, in order to investigate the temperature-to-voltage conversion 
at the local scale within a quantum device \cite{Thomson-1851,McDonald-1962}.  
This probe, which we call Thermoelectric Scanning Gate Microscopy (TSGM), is applied to study the puzzling low-density regime of quantum point contacts (QPC) \cite{Van-Wees-1988,Wharam-1988} .
Our experiments unveils unexpected features in the thermopower that are not visible in conductance measurements.
We explain these deviations by the enhanced sensitivity of the thermopower to phenomena occurring at low transmission.
This observation may help to clarify the nature of the conductance and thermoelectric anomalies in QPCs.


QPCs are quasi-one-dimensional ballistic channels in high-mobility two-dimensional electron gases (2DEG). 
Their conductance curves versus split-gate voltage show quantized plateaus at integer multiples of $2e^2/h$ as a consequence of ballistic transport \cite{Buttiker-1990}. 
They also show anomalous features that are believed to result from electron-electron (e-e) interactions.
The conductance exhibits a shoulder-like feature known as the 0.7 anomaly \cite{Thomas-1996}, which disappears as the temperature is lowered. 
Additionally, the differential conductance exhibits a zero-bias peak at very low temperature, known as the zero-bias anomaly (ZBA) \cite{Cronenwett-2002}.
Many different models have been proposed to explain these anomalies, but after decades of investigations, their exact microscopic origin still 
remains a matter of intense debate \cite{Thomas-1996,Cronenwett-2002,Rejec-2006,Iqbal-2013,Bauer-2013,Brun-2014,Brun-2016,Micolich-2011}. 

The thermoelectric properties of QPCs have also been studied \cite{Streda-1989} and were shown to be excellent probes of quantum confinement effects. 
The Seebeck coefficient $S =(\frac{\partial V}{ \partial T} )_{I=0}$, relates variations of voltage $V$  to temperature $T$ in the absence of current $I$.
It has been shown to oscillate with the number of transmitted modes in the QPC \cite{Molenkamp-1990}. The thermal conductance and Peltier coefficients show similar behavior \cite{Molenkamp-1992}.

In a single-electron picture, $S$ is linked to the conductance $G$ through Mott's relation \cite{Mott-1936}:
\begin{equation}
 S^M (\mu,T) = -\frac{\pi^2 k_B^2 T}{3 e} \ \frac{1}{G(\mu,T)} \ \frac{\partial G(\mu,T)}{\partial \mu}
 \label{Mott},
\end{equation}
where $\mu$ is the chemical potential.
However, the thermopower is predicted to be sensitive to e-e interactions, and often reveals 
different information compared to the conductance. For example, in the case of Coulomb blockade, thermopower measurements 
probe the excitation spectrum rather than the addition spectrum \cite{Beenakker-1992,Dzurak-1997}. 
Since the thermopower is sensitive to the slope of the local density of states (DOS), it may be a useful probe of correlated behavior \cite{Sanchez-2016,Boese-2001}, 
which makes it very relevant in the case of QPC anomalies. 
Large deviations from Eq.(\ref{Mott}) have indeed been reported in QPCs below the first conductance plateau \cite{Appleyard-2000}, 
and were attributed to the important role of e-e interactions in this regime.

\begin{figure}
\includegraphics[scale=0.43]{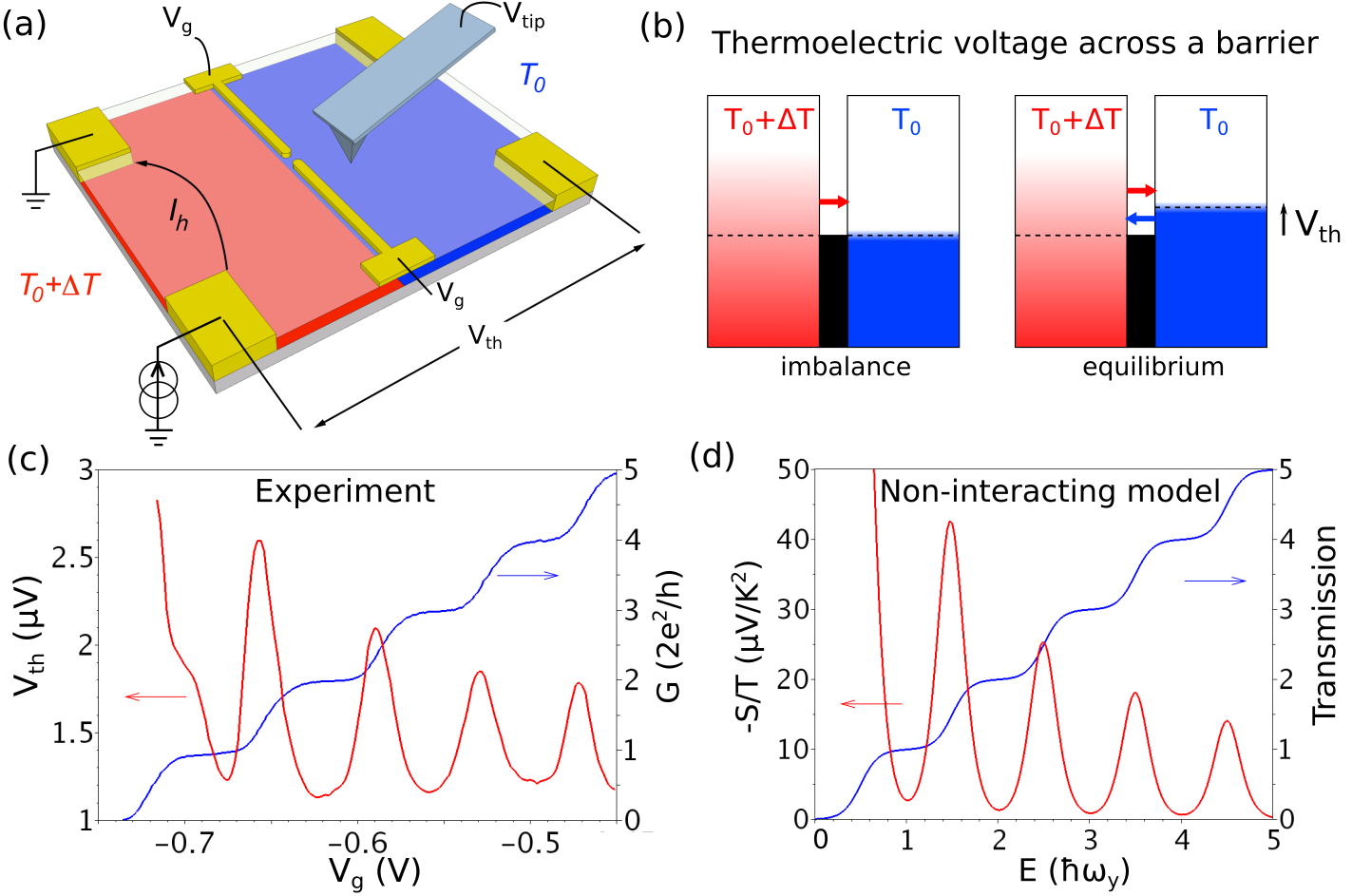}
\caption{\label{fig1}
(a) Scheme of the TSGM experiment: one side of the device is brought to higher temperature and the thermovoltage is recorded while scanning the polarized tip.
(b) Seebeck effect across a barrier (black region): red and blue bars illustrate the energy distribution of charge carriers on the hot and cold sides. 
If transmission is energy-dependent, fluxes of hot and cold carriers are imbalanced. At equilibrium, a charge accumulation on the 
cold side restores a total balance of fluxes: this builds the thermovoltage $V_{th}$.
 (c) Differential conductance $G$, measured using a 4-probes technique at 25 mK (blue) and thermovoltage measured using a heating current of 180 nA (red) versus gate voltage $V_g$. 
(d) Theoretical transmission (blue) and Seebeck coefficient (red) calculated from the saddle-point model \cite{Buttiker-1990} versus energy, for $\omega_x/\omega_y = 0.6$.}
\end{figure}

In this paper, we present an additional perspective on the low-transmission regime of QPCs, 
through interferometric Seebeck measurements performed with our novel TSGM technique.  
This new microscopy mode is a variant of the scanning gate microscopy (SGM) where the negatively polarized tip of a low-temperature scanning probe microscope 
is scanned above the surface, while recording tip-induced changes in the device's conductance \cite{Topinka-2001,Sellier-2011}. 
In TSGM, the device's Seebeck coefficient $S$ is recorded instead of its electrical conductance \cite{Abbout-Thesis}. 
One of the two electron reservoirs is heated using a low frequency AC current
and the thermovoltage $V_{th}$ across the device is measured as a function of the tip position (Fig.\ref{fig1}a). 
The Seebeck coefficient is obtained as $S=V_{th}/\Delta T$ where $\Delta T$ is the temperature difference (Fig.\ref{fig1}b).

In SGM images of QPCs, when the tip voltage is chosen such as to locally deplete the 2DEG, it generates fringes spaced by half of the Fermi wavelength 
due to Fabry-P\'erot interference between the depleted region below the tip and the constriction defined by the split gate \cite{Topinka-2001,Jura-2009,Kozikov-2013}.
These fringes are observed both in SGM and TSGM images, but with significant differences near the QPC pinch-off.
These differences cannot be explained in the framework of Mott's relation (Eq.(\ref{Mott})), indicating the crucial role played by the electron interactions in this regime. 
Understanding this many-body physics has been a research topic for decades, and is beyond the scope of the present paper.

Instead, we analyze in more details the thermoelectric scanning gate interferometry technique in a single-particle framework. 
We propose a microscopic model incorporating the contribution of a resonant energy level located close to the QPC, to simulate the SGM and TSGM signals. 
We show explicitly that the different connection of $G$ and $S$ to the slope of the density of states at the Fermi energy results in an enhanced sensitivity of $S$ to localized states with weak transmission. 
This property makes the TSGM technique particularly relevant for the investigation of the low-density regime of QPCs where spontaneous localized states 
have been predicted \cite{Hirose-2003} and observed \cite{Cronenwett-2002, Morimoto-2003}.

\section{Description of thermopower measurements}

The device chosen to illustrate this new experimental technique is a QPC, defined in a GaAs/AlGaAs heterostructure by a 270-nm-long and 
300-nm-wide gap of a Ti/Au split gate. The 2DEG located 105 nm below the surface has $2.5\times 10^{15}$ m$^{-2}$ electron density
and $1.0\times 10^6$ cm$^2$/(V.s) electronic mobility at low temperature. 
The device is thermally anchored to the mixing chamber of a dilution fridge in front of a cryogenic scanning probe microscope \cite{Hackens-2010}
and cooled down to a base temperature of 25 mK. The four-probe differential conductance is measured by a lock-in technique 
using a 10 $\sf \mu$V excitation, at 77 Hz.
A series resistance of 200 $\sf \Omega$ is subtracted from all data in order to have the first conductance plateau at $2e^2/h$.
The lever-arm parameter of the split gate $\alpha = 54$ meV/V is deduced from the non-linear spectroscopy of the QPC subbands 
separated by $\Delta E = 3.5 $ meV (see Fig.~S1 in the supplemental materials \cite{supplemental}).

\begin{figure}
\includegraphics[scale=0.7]{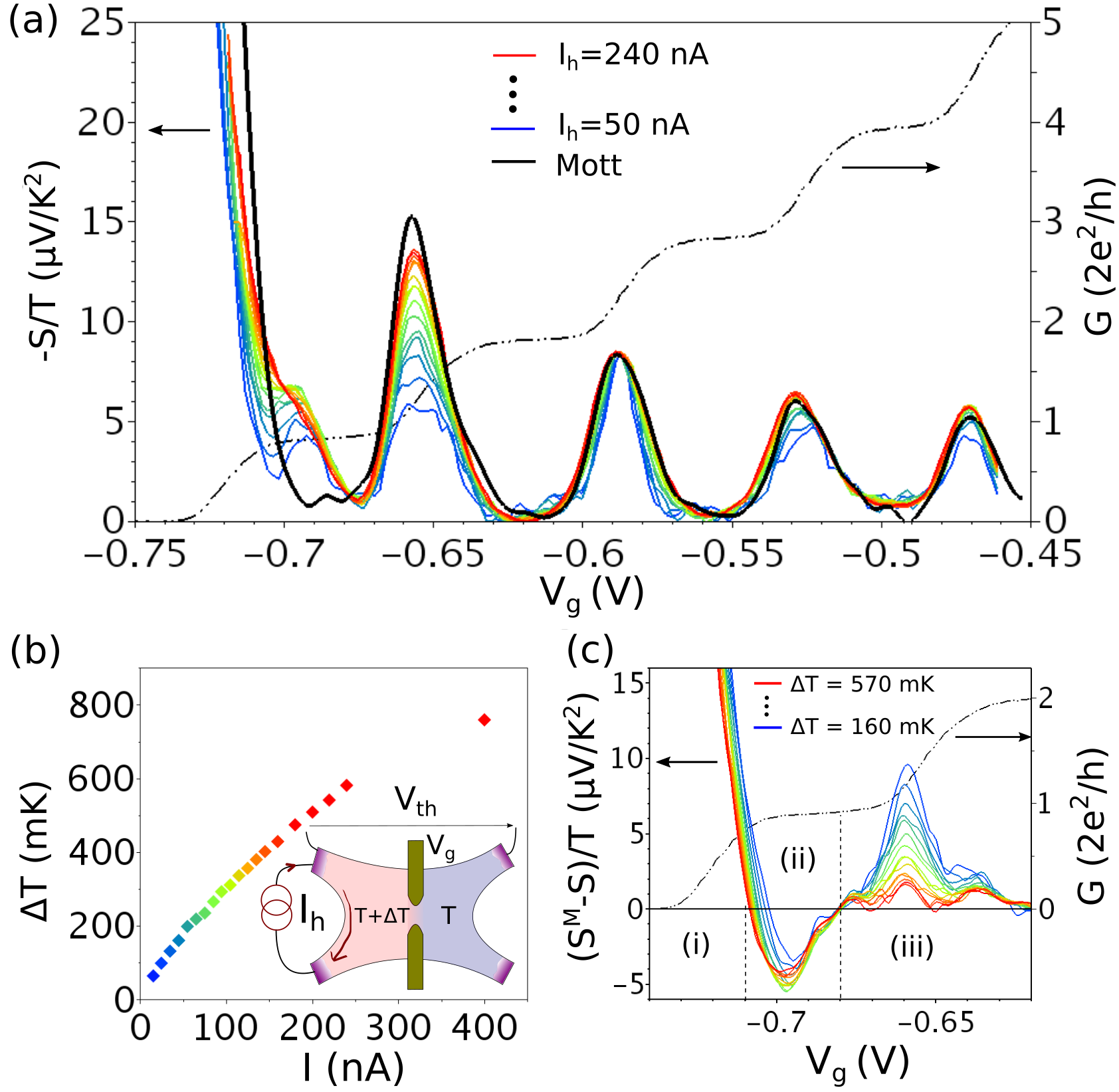}
\caption{\label{fig2} (a) Black curve: $-S^M/T$ calculated using Mott's law from the $G(V_g)$ curve at 25 mK (dashed line). 
Colored curves: $ -S/T = (V_{th}/\Delta T )/T_{\rm average}$ scaled to the black curve at the third peak summit using $\Delta T$ as the only fitting parameter.
Heating currents range from 50 to 240 nA (blue to red). (b) Temperature difference $\Delta T$ as a function of the heating current.
Inset: Electron microscope image of the device. The scale bar is 5 $\mu$m. (c) $ S/T - S^M/T$ for temperature differences from 160 to 570 mK (blue to red) as a function of $V_g$. Same data as in (a).}
\end{figure}

To measure the Seebeck coefficient of the QPC, we use the electron-heating technique depicted in Fig. \ref{fig1}a.
We inject an AC current at 7.17 Hz between two contacts on the same side of the QPC, and record the voltage
across the QPC using a lock-in detection at twice the heating frequency (14.34 Hz), in order to be sensitive to the dissipated power only
and avoid any contribution related to the electrical conductance \cite{Molenkamp-1992}.
The thermovoltage recorded versus gate voltage $V_g$ is shown in Fig. \ref{fig1}c together with the measured conductance curve. 
As expected theoretically\cite{Streda-1989}, the thermovoltage oscillates between minimum values when the QPC is on a plateau and maximum values for transitions between plateaus.
For comparison, the transmission and Seebeck coefficients expected from a non-interacting saddle-point model \cite{Buttiker-1990} are plotted in Fig. \ref{fig1}d. 

A crucial issue in thermopower measurements is to relate the applied heating current to an actual temperature difference $\Delta T$.
To evaluate this quantity, we use two independent methods.
First, we use Mott's law for high densities, assuming that it is valid when more than 2 modes are transmitted through the QPC, and compare quantitatively 
the measured value $S$ with that predicted by Mott's law $S^M$. 
This comparison indicates the existence of a heating-current-dependent but gate-voltage-independent background in the measured signal (see Fig.~S3 in the supplemental materials \cite{supplemental}).
With this background contribution removed, the measured Seebeck coefficient should be given by Mott's relation applied to the measured 
conductance (black curve Fig. \ref{fig2}a). 
This is well verified for the third to fifth transitions, the only fitting parameter being the temperature difference. 

Fig. \ref{fig2}b shows the temperature differences deduced from these assumptions, ranging from 100 to 800 mK for heating currents from 15 to 400 nA.
Second, we compare these extracted values with estimates obtained from the temperature and current dependence of the Shubnikov-de-Haas oscillations 
in our sample, and find a good agreement (see Fig.~S4 in the supplemental materials \cite{supplemental}).
Note that the temperature difference is always larger than the average temperature, such that the system is far from the linear regime.
Nevertheless, it has been shown that Mott's law holds even in this highly non-linear regime, provided 
that $\Delta T$ is smaller than the subbands spacing and smearing \cite{Lunde-2005}.

Interestingly, the electronic temperature in the middle of the heated reservoir evolves sub-linearly with the heating current $I_h$, whereas one could naively
expect a $I_h^2$ dependence related to the dissipated Joule power.
This can be explained by the non-linear temperature dependence of heat losses in 2DEGs, mostly due to phonon emission \cite{Price-1982} and electron out-diffusion 
in the ohmic contacts \cite{Jezouin-2013, Sivre-2017}.
These competing losses yield a non-uniform temperature profile and a sub-linear dependence on $I_h$ of the local temperature far away from the 
ohmic contacts \cite{Mittal-1996}. 

In Fig.\ref{fig2}a, the correspondence between $S$ and $S^M$ does not hold when less than three QPC modes are transmitted, with three distinct features highlighted 
by the plot of their difference (see Fig. \ref{fig2}c):
(i) in the transition from pinch-off to the first plateau ($V_g \sim $-0.74 to -0.71 V), the difference is very large (positive) and increases with lowering temperature, 
(ii) on the first plateau ($V_g \sim $ -0.71 to -0.68 V), the difference is of opposite sign (negative) and forms a peak at the lowest temperatures, 
(iii) in the transition from the first to the second plateau ($V_g \sim $ -0.68 to -0.64 V), a significant difference (positive) arises as $\Delta T$ is lowered below 500 mK.

Differences between $S$ and $S^M$ in the low-density regime of QPCs have already been reported in Ref.~\cite{Appleyard-2000}.
In that work, a local minimum was observed in the thermopower at the position of the 0.7$\times 2e^2/h$ anomalous conductance plateau at 2 K, 
as expected from Eq.(\ref{Mott}). However this minimum was shown to disappear into a shoulder at the lowest temperature of 300 mK, whereas 
the 0.7 plateau was still present (deviation from Eq.(\ref{Mott})). 

Here, the thermopower is also in contradiction with Mott's law but in a different way.  The base temperature is much lower (25 mK) and 
the conductance curve does not show any 0.7 plateau (an absence related to the emergence of a zero-bias peak in the differential conductance \cite{Cronenwett-2002}). 
The thermopower however shows a peak at the lowest temperatures and a minimum around  $V_g = -0.705$~V that disappears for $\Delta T$ above 500 mK (Fig.\ref{fig2}a).

\section{Thermoelectric scanning gate microscopy}

\begin{figure}
\includegraphics[width=1.0\linewidth]{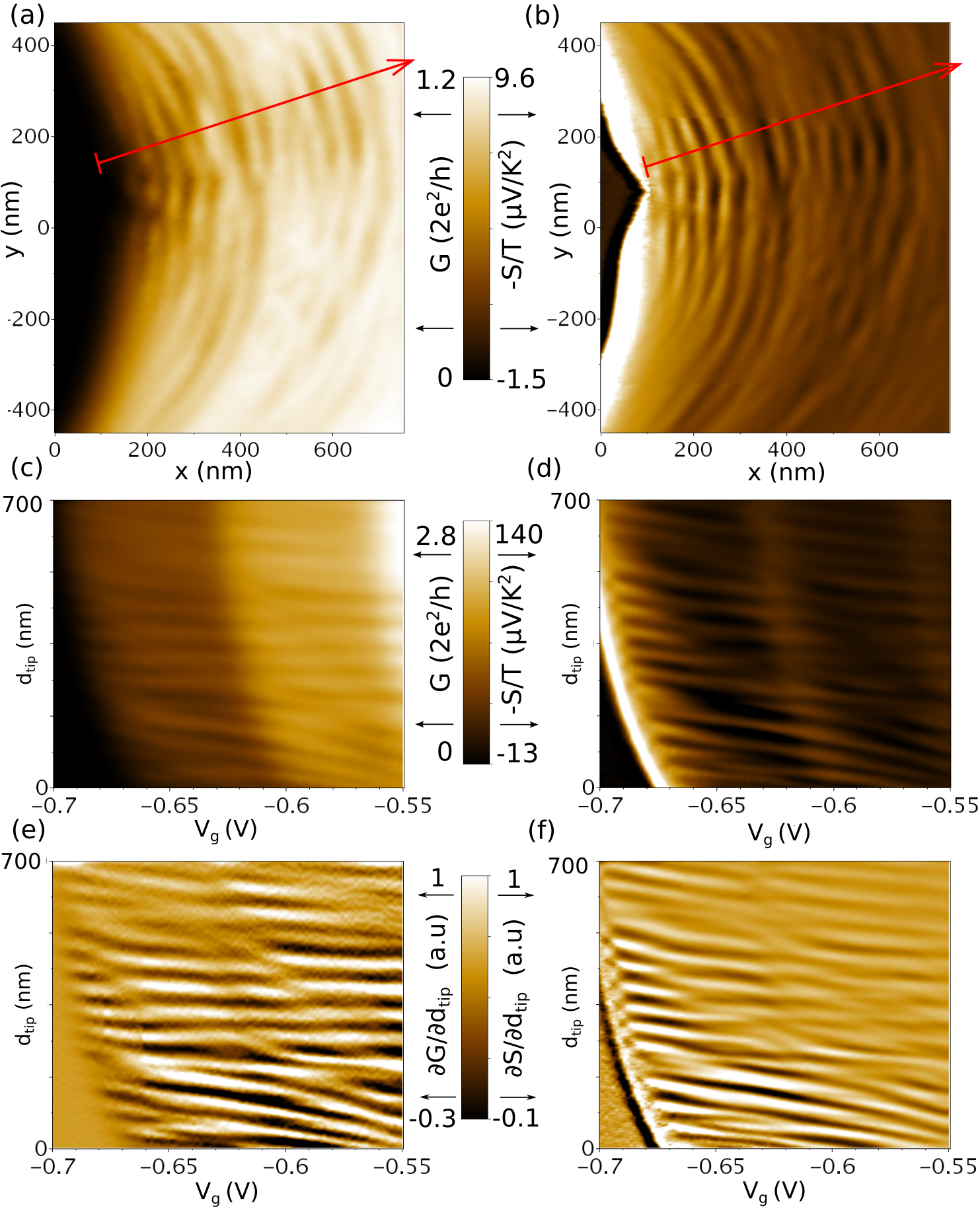}
\caption{\label{fig3} (a) SGM image recorded when the QPC is open on the first plateau ($V_g =$ -0.66 V): conductance $G$ as a function of tip coordinates. 
The QPC center is located at (-150 ; 0) nm on the left side of the image. (b) TSGM image of the thermopower in the same conditions, but recorded in a second 
pass while heating the reservoir on the opposite side of the scanning area.
(c) Conductance and (d) thermovoltage as a function of gate voltage and tip position 
along the red line drawn in (a) and (b), recorded in the exact same conditions, but in separate tip scans.
(e,f) Derivative of (a) and (b) with respect to tip position.}
\end{figure} 

We now report on the TSGM experiment, i.e. the investigation of thermoelectric transport in presence of the scanning gate. The tip is scanned 50 nm above the sample 
surface with an applied voltage of -6V, which locally depletes the 2DEG. Fig. \ref{fig3} shows the SGM and TSGM images obtained by recording successively the conductance and the thermoelectric 
voltage in two different tip scans. The conductance signal is recorded using an AC excitation of $10\ \rm \mu$V,
and the thermovoltage using a heating current of 150 nA, corresponding to a temperature difference of 450 mK.
Both images look very similar, presenting interference fringes due to the Fabry-P\'erot cavity formed by the QPC and the tip-depleted region. 
The TSGM image provides to our knowledge the first observation in real space of a thermally-driven electron interferometer.

Despite their apparent similarities, these images carry distinct information and one cannot be deduced from the other.
Indeed, even in the range where Mott's relation (\ref{Mott}) is valid, deducing the TSGM image $S(x_{tip},y_{tip})$ from the SGM $G(x_{tip},y_{tip})$
would require a knowledge of how the transmission evolves with the chemical potential ($\partial G/\partial \mu$). 
Most of the time, this quantity is not available in GaAs 2DEG, since it requires a backgate to vary the global electron density\cite{Braem-2018,Braem-2018b},
which is a real challenge in high-mobility GaAs heterostructures.

In Fig.\ref{fig2}a, the chemical potential of the QPC itself was identified as being proportional to $V_g$, which is a reasonable approximation
when considering only the QPC transmission. In TSGM images, the distant influence of the tip does not allow such an identification, since 
the studied system now consists of the QPC coherently coupled to the tip-induced Fabry-P\'erot cavity. 
As a consequence, the chemical potential of the system is not linearly linked to $V_g$, and TSGM images provide a different information than what is obtained from SGM.

To illustrate these differences, we study the evolution of the interference fringes as a function of the QPC opening.
The conductance $G$ and the thermo-voltage $V_{th}$ are recorded separately while scanning the tip along the line shown in Fig. \ref{fig3}(a,b), using 
an excitation of 15 $ \mu$V and a temperature difference of 450 mK for $G$ and $S$ measurements, respectively.
The evolution of the interference fringes with $V_g$ is shown in Fig.\ref{fig3}(c,d), for the conductance and thermovoltage signals. 
The fringes are similar in both signals but are superimposed on two very different background signals : a series of 
conductance plateaus for $G$, and a series of peaks for $V_{th}$ (see Fig.\ref{fig2}), including a very strong peak at the QPC pinch-off.
To highlight the fringes' evolution with $V_g$, we plot the first derivative of both signals with respect to the tip position
along the red line, $d_{tip}$, and plot the results in Fig.\ref{fig3}(e,f).
In these maps, we observe a complex evolution of the interference fringes, which looks globally similar in $G$ and $S$, 
though few differences can be detected.

\begin{figure}
\includegraphics[scale=0.53]{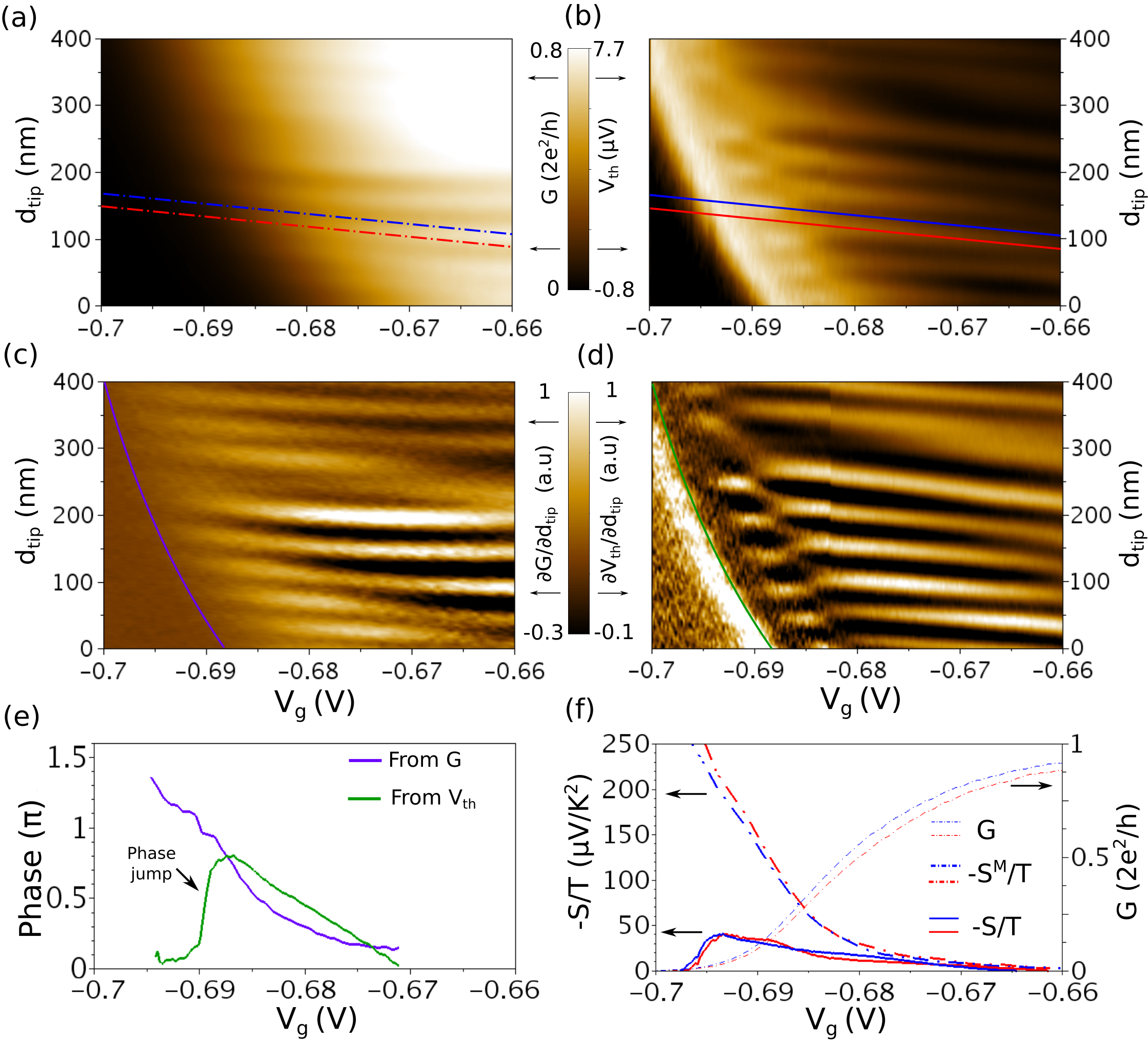}
\caption{\label{fig4} (a) Conductance and (b) thermovoltage as a function of gate voltage and tip position 
along the red line drawn in Fig. \ref{fig3}, where $d_{tip}$ here denotes tip position between 250 and 650 nm in Fig.\ref{fig3}c-f.
(c,d) Derivative of (a) and (b) with respect to tip position. (e) Phase of the interference fringes as a function of gate voltage extracted 
from the conductance (purple) and thermovoltage (green), following the pinch-off lines shown in (c) and (d) to account for cross-talk effect. 
(f) Line profiles of $G$, $-S^M/T$ and $-S/T$ extracted along the red and blue lines in (a) and (b).
}
\end{figure}

In the following, we focus on the low conductance regime, below $\sim 0.5\times 2e^2/h$, where $S$ and $S^M$ differ by a large amount (Fig. \ref{fig2}c).
In this regime, the SGM and TSGM interferometric signals evolve differently with gate voltage.
The conductance oscillations follow a monotonic behavior (Fig.\ref{fig4}a,c), i.e.\@ their phase evolves monotonically with gate voltage, 
whereas the thermopower oscillations exhibit an abrupt phase shift at a conductance of about $0.25 \times 2e^2/h$ (Fig. \ref{fig4}b,d), which can also be surmised in Fig.\ref{fig3}d,f. 
A Fourier analysis (Fig. \ref{fig4}e) indicates that the phase shift observed in the thermopower is almost $\pi$. 
Similar phase shifts have been observed in the conductance signal in many different devices during our previous experiments \cite{Brun-thesis}.
However they were observed at higher transmission, close to the first conductance plateau.
Here the phase shift is observed in the thermopower at very low transmission, where no phase shift is present in the conductance.

\section{Discussion}

Phase shifts in the conductance were previously observed in Aharonov-Bohm interferometers
containing a quantum dot in one arm, where the interference pattern experiences 
a phase shift by $\pi$ whenever one charge is added to the quantum dot \cite{Yacoby-1995}, and by $\pi/2$ in the Kondo regime \cite{Zaffalon-2008}.

Similarly, the phase shift observed in our SGM-based interferometry experiment indicates the presence of a resonant level in the cavity formed by the QPC and the tip. 
Such a resonant level could be located in the QPC itself, as observed in Ref.~\cite{Brun-2016}, or in the 2DEG region between the QPC and the tip, 
as a result of potential fluctuations induced by remote ionized dopants. In the latter case, the phase shift of the interference pattern would barely depend on the QPC gate voltage. 
Experimentally however, the phase shift evolves with gate voltage following exactly the QPC pinch-off line (Fig.\ref{fig4}d), suggesting that this resonant state is in the close vicinity of the QPC. 

Alhough this localized state could result from disorder in the potential of the QPC channel, it could also correspond to a spontaneously localized charge.
Indeed, several evidences for the existence of bound states in QPCs near pinch-off have been reported \cite{Morimoto-2003, Puller-2004, Steinberg-2006,Yoon-2007, Wu-2012,Ho-2018}.
Such bounds states have been predicted to spontaneously form in QPCs by several numerical studies \cite{Sushkov-2001,Sushkov-2003, Shulenburger-2008,Soffing-2009,Guclu-2009},
as a consequence of e-e interactions.
When the potential barrier of the QPC is above the Fermi level, there are two regions of low density, one on each side of the barrier, 
where charges could spontaneously localize. 
Numerical simulations of this peculiar situation encountered near pinch off supports the presence of
bound states \cite{Berggren-2008,Yakimenko-2013} detected in coupled QPCs experiments \cite{Yoon-2007, Yoon-2009}.
This scenario has also been proposed in Ref.\cite{Ren-2010} to explain the presence of QPC conductance anomalies down to very low conductance.
This is also consistent for example with the results of local spin density functional theory presented in Ref.\cite{Rejec-2006}, where
two charges are shown to be localized on both sides of the main barrier at low density.
Finally, classical electrostatic simulations also confirm that at low QPC transmission, two one-dimensional regions form on both sides of the channel, 
where the density is low enough to induce Wigner crystallization \cite{Wigner-1934} (see section VI in the supplemental materials \cite{supplemental}).

It should be mentionned, however, that some experiments \cite{Kawamura-2015} found no sign of such a localized state, and that some theoretical works proposed an alternative 
explanation for the related 0.7 and zero-bias anomalies \cite{Bauer-2013, Schimmel-2017}, whithout invoking the presence of a localized charge.
Note also that, in presence of interactions, Friedel oscillations between the QPC barrier and the SGM tip \cite{Freyn-2008, Soffing-2009} could give rise to a phase shift of the 
interference pattern, but this effect is not expected to appear as abruptly versus gate voltage as it is observed here.

In conclusion, the phase shift observed here by TSGM at very-low transmission is probably related to the presence of a localized state on the side of the QPC, 
but distinguishing whereas this state is induced by e-e interaction or by disorder is beyond the scope of the present paper and would require additional investigation.

\section{ Effect of a localized state in TSGM interferometry}

\begin{figure}[h!]
\includegraphics[width = 1.0\linewidth]{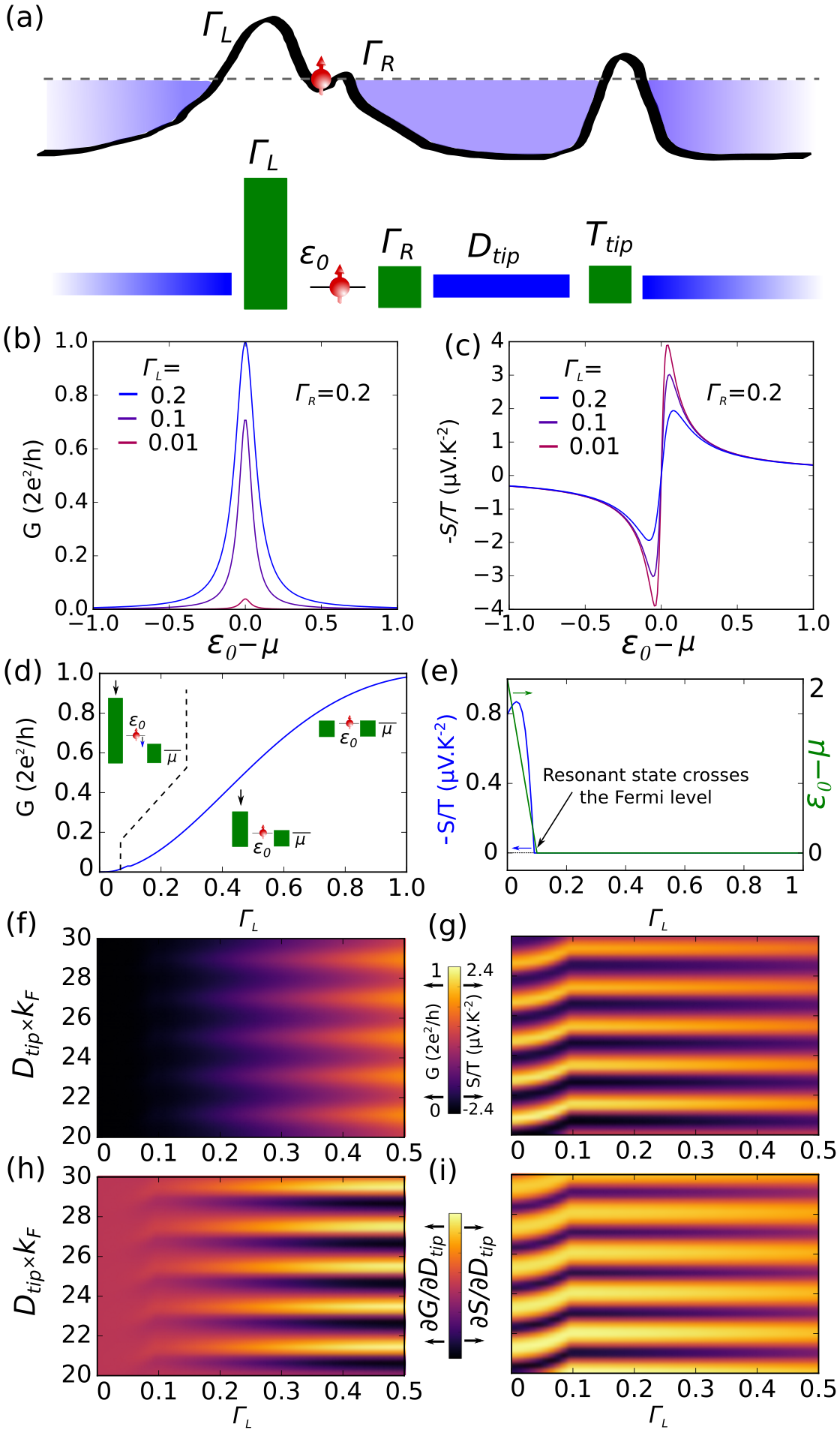}
\caption{\label{fig5} (a) Up: typical energetic potential which could lead to the proposed scenario: a localized state is located on the tip side
of the main QPC barrier. Down: Scheme of the 1D model including, from left to right, the QPC barrier ($\Gamma_L$), the localized state ($\epsilon_0$ and $\Gamma_R$), 
and the tip-induced cavity ($D_{tip}$ and $T_{tip}$). (b) Conductance of the resonant level as a function of its energy relative to $\mu$ for fixed $\Gamma_R = 0.2$ and 
$\Gamma_L$ = 0.2 (blue) to 0.01 (red).(c) Seebeck coefficient in the same configuration. (d) Conductance evolution with $\Gamma_L$ assuming that the resonant level, perfectly coupled to the 
right lead ($\Gamma_R$ = 1), crosses $\mu$ at $\Gamma_L$ = 0.1 and then stays just below the Fermi level while the QPC opens. (e) Seebeck coefficient in the exact same scenario.
(f,g): Evolution of the Fabry-P\'erot interference in this scenario: conductance (f) and Seebeck coefficient (g) as a function of tip distance and $\Gamma_L$.  
(h,i): Derivative of $G$ and $S$ with respect to $D_{tip}$.
}
\end{figure}

In this section, we use a simple microscopic model to analyze the effect of a localized state (irrespectively of its origin in this specific sample),
on the SGM and TSGM interferometric signals.
We model the gate-controlled potential in the QPC by a barrier with transmission rate $\Gamma_L$ and we assume the presence of a 
localized state on the right side of the barrier, at an energy $\epsilon_0$, separated from the right 
reservoir by a barrier with transmission rate $\Gamma_R$ (Fig.\ref{fig5}a).
We also model the tip as a distant scatterer of transmission amplitude $T_{tip} = 0.99$, whose distance from the resonant level $D_{tip}$ can be varied. 
The expression of the energy-dependent transmission $T(E)$ through the whole system can then be calculated exactly as presented in Ref.\cite{Brun-2016}.
In the Landauer framework, $G$ and $S$ respectively express as \cite{Note1}:
\begin{eqnarray}
G = \frac{2e^2 }{h} L_0 \\
S = -\frac{1}{\rvert e\rvert T} \frac{L_1}{L_0}
\end{eqnarray}

where 
\begin{eqnarray}
 L_m&=&\int_{-\infty}^{+\infty} (E -\mu )^m \left( \frac{\partial f}{\partial E} \right) T(E) \, \mathrm{d}E, 
\end{eqnarray}
and $f(E)$ is the Fermi distribution.

We first show that the behaviors with respect to the tunneling rates of asymmetric barriers are very different for $G$ and $S$ (Fig.\ref{fig5}b-c).
For a fixed tunneling rate $\Gamma_R$ = 0.2 (in units of the hopping term $t$), the effect of a decrease in $\Gamma_L$ is opposite for $G$ and $S$: 
it tends to decrease $G$, which is dominated by the lowest tunneling rate, 
but to increase $S$ on both sides of the resonance since it is sensitive to the resonance sharpness, and thus inversely proportional to $\Gamma_{tot}=\Gamma_L+\Gamma_R$.

In the following, we assume that the resonant state is very well coupled to the right lead ($\Gamma_R $ fixed at 1) and that it evolves with $\Gamma_L$ as shown in the inset of Fig.\ref{fig5}e.
As the QPC is progressively opened from the pinch-off (increasing of $\Gamma_L$), the energy of the resonant state $\epsilon_0$ drops until it reaches 
the Fermi level at $\Gamma_L = 0.1$ (dashed line in Fig.\ref{fig5}d). 
Opening the QPC further does not change $\epsilon_0$, but the level stays pinned close to $\mu$, up to the point where $\Gamma_L = \Gamma_R$ and the transmission reaches unity.  
In this scenario, the conductance behaves as normally expected for a QPC (Fig.\ref{fig5}d): the resonance 
is almost invisible as it appears close to the pinch-off. It also gives no discernible signature in the interference fringes, since it occurs at very low transmission (Fig.\ref{fig5}f and \ref{fig5}h).  
The high sensitivity of $S$ results from the strong variation of the transmission with energy when the resonance approaches the Fermi level.

The phase shift induced by the drop of $\epsilon_0$ below $\mu$ is highly visible in $S$, and is due to the Breit-Wigner like resonance of the localized state,
that induces a total $\pi$ phase shift when the level crosses the Fermi energy.
Since the Fabry-P\'erot cavity probes twice the resonant level phase shift in case of a strong coupling on the cavity side\cite{Brun-2016}, the total
phase shift should be $2\pi$. However assuming that the $\epsilon_0$ level falls just below $\mu$ but remains at a constant energy value,
the phase shift remains half the value of  $2\pi$, which corresponds to the experimentally observed $\pi$ value. 
This model therefore provides a plausible scenario to understand our TSGM interferometry results

\section{Conclusion}

In conclusion, we introduced a new scanning probe technique to image thermoelectric transport through a QPC. 
By scanning the polarized tip in front of the QPC, we imaged for the first time interference of electrons driven by a temperature difference, 
in analogy with the well-established SGM experiments where the electron flow is driven by a voltage difference. 
In addition, we showed that in the very-low-conductance regime, the thermopower interference fringes experience an abrupt phase shift, invisible in the conductance signal. 
We propose a simple 1D model to show that this phase shift and its characteristics can be explained by the contribution of a localized
state inside or close to the QPC, and that stays pinned to the Fermi energy of the leads as the first QPC mode opens.
The fact that this localized state signature is hidden in conductance measurements but highly visible in the Seebeck coefficient
is explained by its sharpness and occurrence at low transmission.
This work illustrates that the combination of scanning gate microscopy and thermoelectric measurements can unveil elusive 
phenomena that escape transport measurements.
Though we cannot draw definitive conclusions on the mechanism leading to this localization, we provide a new tool that may prove useful in
future investigations of conductance and thermoelectric anomalies in QPCs.

%

We thank J.-L. Pichard and A. Abbout for the original idea of the experiment.
This work was supported by the ``Cr\'edit de Recherche'' (CDR) grant no. J.0009.16 from the FRS-FNRS, and
the French Agence Nationale de la Recherche (``ITEM-exp''project). 
B.B, F.M. and B.H. acknowledge support from the Belgian FRS-FNRS, S.F. received support from the FSR at
UCL.

%
%

\begin{thebibliography}{64}%

\bibitem{Dresselhaus-2007}
M.~Dresselhaus, G.~Chen, M.~Tang, R.~Yang, H.~Lee, D.~Wang, Z.~Ren, J.-P.
  Fleurial, and P.~Gogna, \emph{New Directions for Low-Dimensional
  Thermoelectric Materials}, 
  \href{http://dx.doi.org/10.1002/adma.200600527}{Adv. Mat. \textbf{19}, 1043--1053 (2007).}

\bibitem{Jiang-2016}
J.-H. Jiang and Y.~Imry, \emph{Linear and nonlinear mesoscopic thermoelectric
  transport with coupling with heat baths},
  \href{http://www.sciencedirect.com/science/article/pii/S1631070516300858}
  {Comptes Rendus Physique \textbf{17}, 1047 - 1059 (2016).}

\bibitem{Giazotto-2012}
F.~Giazotto and M.~J. Mart{\'i}nez-P{\'e}rez, \emph{The Josephson heat
  interferometer}, \href{http://dx.doi.org/10.1038/nature11702}
  {Nature \textbf{492}, 401 (2012).}

\bibitem{Jezouin-2013}
S.~Jezouin, F.~D. Parmentier, A.~Anthore, U.~Gennser, A.~Cavanna, Y.~Jin, and
  F.~Pierre, \emph{Quantum Limit of Heat Flow Across a Single Electronic
  Channel}, \href{http://science.sciencemag.org/content/342/6158/601}{
  Science \textbf{342}, 601--604 (2013).}

\bibitem{Cui-2017}
L.~Cui, W.~Jeong, S.~Hur, M.~Matt, J.~C. Kl{\"o}ckner, F.~Pauly, P.~Nielaba,
  J.~C. Cuevas, E.~Meyhofer, and P.~Reddy, \emph{Quantized thermal transport in
  single-atom junctions},\href{http://science.sciencemag.org/content/355/6330/1192}{
  Science \textbf{355}, 1192--1195 (2017).}

\bibitem{Banerjee-2017}
M.~Banerjee, M.~Heiblum, A.~Rosenblatt, Y.~Oreg, D.~E. Feldman, A.~Stern, and
  V.~Umansky, \emph{Observed quantization of anyonic heat flow}, 
  \href{http://dx.doi.org/10.1038/nature22052}{Nature
  \textbf{545}, 75 (2017).}

\bibitem{Sivre-2017}
E.~Sivre, A.~Anthore, F.~D. Parmentier, A.~Cavanna, U.~Gennser, A.~Ouerghi,
  Y.~Jin, and F.~Pierre, \emph{Heat Coulomb blockade of one ballistic channel},
 \href{http://dx.doi.org/10.1038/nphys4280}{ Nat. Phys. \textbf{14}, 145
  (2017).}

\bibitem{Dutta-2017}
B.~Dutta, J.~T. Peltonen, D.~S. Antonenko, M.~Meschke, M.~A. Skvortsov,
  B.~Kubala, J.~K\"onig, C.~B. Winkelmann, H.~Courtois, and J.~P. Pekola,
  \emph{Thermal Conductance of a Single-Electron Transistor}, 
  \href{https://link.aps.org/doi/10.1103/PhysRevLett.119.077701}{Phys. Rev. Lett.
  \textbf{119}, 077701 (2017).}

\bibitem{Halbertal2016}
D.~Halbertal, J.~Cuppens, M.~B. Shalom, L.~Embon, N.~Shadmi, Y.~Anahory, H.~R.
  Naren, J.~Sarkar, A.~Uri, Y.~Ronen, Y.~Myasoedov, L.~S. Levitov,
  E.~Joselevich, A.~K. Geim, and E.~Zeldov, \emph{Nanoscale thermal imaging of
  dissipation in quantum systems},\href{http://dx.doi.org/10.1038/nature19843}{
  Nature \textbf{539}, 407-410 (2016).}

\bibitem{Halbertall-2017}
D.~Halbertal, M.~Ben~Shalom, A.~Uri, K.~Bagani, A.~Y. Meltzer, I.~Marcus,
  Y.~Myasoedov, J.~Birkbeck, L.~S. Levitov, A.~K. Geim, and E.~Zeldov,
  \emph{Imaging resonant dissipation from individual atomic defects in
  graphene},\href{http://science.sciencemag.org/content/358/6368/1303}{
  Science \textbf{358}, 1303--1306 (2017).}

\bibitem{Thomson-1851}
W.~Thomson, \emph{On a mechanical theory of thermoelectric currents.}, Proc.
  Roy. Soc. Edinburgh, 91 - 98 (1851).

\bibitem{McDonald-1962}
D.~K. C.~M. Donald, \emph{Thermoelectricity: An Introduction to the
  Principles}, John Wiley and Sons, Inc., New York, 1962.

\bibitem{Van-Wees-1988}
B.~J. van Wees, H.~van Houten, C.~W.~J. Beenakker, J.~G. Williamson, L.~P.
  Kouwenhoven, D.~van~der Marel, and C.~T. Foxon, \emph{Quantized conductance
  of point contacts in a two-dimensional electron gas}, 
  \href{https://journals.aps.org/prl/abstract/10.1103/PhysRevLett.60.848}{Phys. Rev. Lett.
  \textbf{60}, 848--850 (1988).}

\bibitem{Wharam-1988}
D.~A. Wharam, T.~J. Thornton, R.~Newbury, M.~Pepper, H.~Ahmed, J.~E.~F. Frost,
  D.~G. Hasko, D.~C. Peacock, D.~A. Ritchie, and G.~A.~C. Jones,
  \emph{One-dimensional transport and the quantisation of the ballistic
  resistance},\href{http://stacks.iop.org/0022-3719/21/i=8/a=002}{
  Journal of Physics C: Solid State Physics \textbf{21}, L209 (1988).}

\bibitem{Buttiker-1990}
M.~B\"uttiker, \emph{Quantized transmission of a saddle-point constriction},
  \href{https://journals.aps.org/prb/abstract/10.1103/PhysRevB.41.7906}{Phys. Rev. B \textbf{41}, 7906--7909 (1990).}

\bibitem{Thomas-1996}
K.~J. Thomas, J.~T. Nicholls, M.~Y. Simmons, M.~Pepper, D.~R. Mace, and D.~A.
  Ritchie, \emph{Possible Spin Polarization in a One-Dimensional Electron Gas},
 \href{https://journals.aps.org/prl/abstract/10.1103/PhysRevLett.77.135}{ Phys. Rev. Lett. \textbf{77}, 135-138 (1996).}

\bibitem{Cronenwett-2002}
S.~M. Cronenwett, H.~J. Lynch, D.~Goldhaber-Gordon, L.~P. Kouwenhoven, C.~M.
  Marcus, K.~Hirose, N.~S. Wingreen, and V.~Umansky, \emph{Low-Temperature Fate
  of the 0.7 Structure in a Point Contact: A Kondo-like Correlated State in an
  Open System},\href{https://journals.aps.org/prl/abstract/10.1103/PhysRevLett.88.226805}
  { Phys. Rev. Lett. \textbf{88}, 226805 (2002).}

\bibitem{Rejec-2006}
T.~Rejec and Y.~Meir, \emph{Magnetic impurity formation in quantum point
  contacts}, \href{https://www.nature.com/articles/nature05054}{Nature \textbf{442}, 900--903 (2006).}

\bibitem{Iqbal-2013}
M.~J. Iqbal, R.~Levy, E.~J. Koop, J.~B. Dekker, J.~P. de~Jong, J.~H.~M. van~der
  Velde, D.~Reuter, A.~D. Wieck, R.~Aguado, Y.~Meir, and C.~H. van~der Wal,
  \emph{Odd and even Kondo effects from emergent localization in quantum point
  contacts}, \href{http://dx.doi.org/10.1038/nature12491}{Nature \textbf{501}, 79--83
  (2013).}

\bibitem{Bauer-2013}
F.~Bauer, J.~Heyder, E.~Schubert, D.~Borowsky, D.~Taubert, B.~Bruognolo,
  D.~Schuh, W.~Wegscheider, J.~von Delft, and S.~Ludwig, \emph{Microscopic
  origin of the 0.7-anomaly in quantum point contacts},
  \href{https://www.nature.com/articles/nature12421}
  {Nature \textbf{501}, 73--78 (2013).}

\bibitem{Brun-2014}
B.~Brun, F.~Martins, S.~Faniel, B.~Hackens, G.~Bachelier, A.~Cavanna,
  C.~Ulysse, A.~Ouerghi, U.~Gennser, D.~Mailly, S.~Huant, V.~Bayot, M.~Sanquer,
  and H.~Sellier, \emph{Wigner and Kondo physics in quantum point contacts
  revealed by scanning gate microscopy},
  \href{http://dx.doi.org/10.1038/ncomms5290}{Nat. Commun. \textbf{5}, 4290
  (2014).}

\bibitem{Brun-2016}
B.~Brun, F.~Martins, S.~Faniel, B.~Hackens, A.~Cavanna, C.~Ulysse, A.~Ouerghi,
  U.~Gennser, D.~Mailly, P.~Simon, S.~Huant, V.~Bayot, M.~Sanquer, and
  H.~Sellier, \emph{Electron Phase Shift at the Zero-Bias Anomaly of Quantum
  Point Contacts}, \href{https://link.aps.org/doi/10.1103/PhysRevLett.116.136801}{
  Phys. Rev. Lett. \textbf{116}, 136801 (2016).}

\bibitem{Micolich-2011}
A.~P. Micolich, \emph{What lurks below the last plateau: experimental studies
  of the 0.7 $\times 2e^2 / h$ conductance anomaly in one-dimensional systems},
  Journal of Physics: Cond. Matt. \textbf{23}, 443201 (2011).

\bibitem{Streda-1989}
P.~Streda, \emph{Quantised thermopower of a channel in the ballistic regime},
  \href{http://stacks.iop.org/0953-8984/1/i=5/a=021}{ Journal of Physics: Cond. Matt. \textbf{1}, 1025
  (1989).}

\bibitem{Molenkamp-1990}
L.~W. Molenkamp, H.~van Houten, C.~W.~J. Beenakker, R.~Eppenga, and C.~T.
  Foxon, \emph{Quantum oscillations in the transverse voltage of a channel in
  the nonlinear transport regime},\href{http://link.aps.org/doi/10.1103/PhysRevLett.65.1052}{
  Phys. Rev. Lett. \textbf{65}, 1052--1055 (1990).}

\bibitem{Molenkamp-1992}
L.~W. Molenkamp, T.~Gravier, H.~van Houten, O.~J.~A. Buijk, M.~A.~A. Mabesoone,
  and C.~T. Foxon, \emph{Peltier coefficient and thermal conductance of a
  quantum point contact},\href{https://link.aps.org/doi/10.1103/PhysRevLett.68.3765}{
  Phys. Rev. Lett. \textbf{68}, 3765--3768 (1992).}

\bibitem{Mott-1936}
H.~Mott, N.F.~Jones, \emph{The Theory Of The Properties Of Metals And Alloys},
  Oxford University Press, 1936.

\bibitem{Beenakker-1992}
C.~W.~J. Beenakker and A.~A.~M. Staring, \emph{Theory of the thermopower of a
  quantum dot}, \href{https://link.aps.org/doi/10.1103/PhysRevB.46.9667}{
  Phys. Rev. B \textbf{46}, 9667--9676 (1992).}

\bibitem{Dzurak-1997}
A.~S. Dzurak, C.~G. Smith, C.~H.~W. Barnes, M.~Pepper, L.~Mart\'{\i}n-Moreno,
  C.~T. Liang, D.~A. Ritchie, and G.~A.~C. Jones, \emph{Thermoelectric
  signature of the excitation spectrum of a quantum dot}, 
  \href{https://link.aps.org/doi/10.1103/PhysRevB.55.R10197}{Phys. Rev. B
  \textbf{55}, R10197--R10200 (1997).}

\bibitem{Sanchez-2016}
D.~Sanchez and R.~Lopez, \emph{Nonlinear phenomena in quantum thermoelectrics
  and heat},\href{http://www.sciencedirect.com/science/article/pii/S1631070516300846}{
  Comptes Rendus de Physique \textbf{17}, 1060 - 1071 (2016).}

\bibitem{Boese-2001}
D.~Boese and R.~Fazio, \emph{Thermoelectric effects in Kondo-correlated quantum
  dots},\href{http://stacks.iop.org/0295-5075/56/i=4/a=576}{ Europhys. Lett. \textbf{56}, 576
  (2001).}

\bibitem{Appleyard-2000}
N.~J. Appleyard, J.~T. Nicholls, M.~Pepper, W.~R. Tribe, M.~Y. Simmons, and
  D.~A. Ritchie, \emph{Direction-resolved transport and possible many-body
  effects in one-dimensional thermopower}, 
  \href{https://journals.aps.org/prb/abstract/10.1103/PhysRevB.62.R16275}
  {Phys. Rev. B \textbf{62}, R16275--R16278 (2000).}

\bibitem{Topinka-2001}
M.~A. Topinka, B.~J. LeRoy, R.~M. Westervelt, S.~E.~J. Shaw, R.~Fleischmann,
  E.~J. Heller, K.~D. Maranowski, and A.~C. Gossard, \emph{Coherent branched
  flow in a two-dimensional electron gas}, 
  \href{https://www.nature.com/articles/nphys756}{Nature \textbf{410}, 183--186 (2001).}

\bibitem{Sellier-2011}
H.~Sellier, B.~Hackens, M.~G. Pala, F.~Martins, S.~Baltazar, X.~Wallart,
  L.~Desplanque, V.~Bayot, and S.~Huant, \emph{On the imaging of electron
  transport in semiconductor quantum structures by scanning-gate microscopy:
  successes and limitations}, \href{http://stacks.iop.org/0268-1242/26/i=6/a=064008}{
  Semiconductor Science and Technology \textbf{26}, 064008 (2011).}

\bibitem{Abbout-Thesis}
A.~Abbout, \emph{Thermoelectric transport in quantum point contacts and chaotic
  cavities : thermal effects and fluctuations}, 
  \href{http://tel.archives-ouvertes.fr/tel-00793816}{Ph.{D}. thesis, Universit\'e
  Paris VI (2011).}

\bibitem{Jura-2009}
M.~P. Jura, M.~A. Topinka, M.~Grobis, L.~N. Pfeiffer, K.~W. West, and
  D.~Goldhaber-Gordon, \emph{Electron interferometer formed with a scanning
  probe tip and quantum point contact},
  \href{http://link.aps.org/doi/10.1103/PhysRevB.80.041303}
  {Phys. Rev. B \textbf{80}, 041303 (2009).}

\bibitem{Kozikov-2013}
A.~A. Kozikov, C.~R{\"o}ssler, T.~Ihn, K.~Ensslin, C.~Reichl, and W.~Wegscheider,
  \emph{Interference of electrons in backscattering through a quantum point
  contact}, New Journal of Physics \textbf{15}, 013056 (2013).

\bibitem{Hirose-2003}
K.~Hirose, Y.~Meir, and N.~S. Wingreen, \emph{Local Moment Formation in Quantum
  Point Contacts}, \href{https://link.aps.org/doi/10.1103/PhysRevLett.90.026804}{
  Phys. Rev. Lett. \textbf{90}, 026804 (2003).}

\bibitem{Morimoto-2003}
T.~Morimoto, Y.~Iwase, N.~Aoki, T.~Sasaki, Y.~Ochiai, A.~Shailos, J.~P. Bird,
  M.~P. Lilly, J.~L. Reno, and J.~A. Simmons, \emph{Nonlocal resonant
  interaction between coupled quantum wires}, 
  \href{https://doi.org/10.1063/1.1579851}{ Applied Physics Letters
  \textbf{82}, 3952-3954 (2003).}

\bibitem{Hackens-2010}
B.~Hackens, F.~Martins, S.~Faniel, C.~A. Dutu, H.~Sellier, S.~Huant, M.~Pala,
  L.~Desplanque, X.~Wallart, and V.~Bayot, \emph{Imaging Coulomb islands in a
  quantum Hall interferometer},\href{https://www.nature.com/articles/ncomms1038}
  {Nat. Commun. \textbf{1}, 39 (2010).}

\bibitem{supplemental}
\emph{See supplemental materials for additional data and analysis, which also
  includes Refs.\cite{Cronenwett-thesis, Iqbal-thesis, Karavolas-1991,
  Fletcher-1995, Schmidt-2012, Tanatar-1989}}.

\bibitem{Lunde-2005}
A.~M. Lunde and K.~Flensberg, \emph{On the Mott formula for the thermopower of
  non-interacting electrons in quantum point contacts}, 
  \href{http://stacks.iop.org/0953-8984/17/i=25/a=014}{ Journal of Physics:
  Condensed Matter \textbf{17}, 3879 (2005).}

\bibitem{Price-1982}
P.~J. Price, \emph{Hot electrons in a GaAs heterolayer at low temperature},
 \href{https://doi.org/10.1063/1.330026}{
 Journal of Applied Physics \textbf{53}, 6863-6866 (1982).}

\bibitem{Mittal-1996}
A.~Mittal, R.~Wheeler, M.~Keller, D.~Prober, and R.~Sacks,
  \emph{Electron-phonon scattering rates in GaAs/AlGaAs 2DEG samples below 0.5
  K}, \href{http://www.sciencedirect.com/science/article/pii/0039602896004645}{
  Surface Science \textbf{361}, 537 - 541 (1996).}

\bibitem{Braem-2018}
B.~A. Braem, C.~Gold, S.~Hennel, M.~R{\"o}{\"o}sli, M.~Berl, W.~Dietsche,
  W.~Wegscheider, K.~Ensslin, and T.~Ihn, \emph{Stable branched electron flow},
 \href{http://stacks.iop.org/1367-2630/20/i=7/a=073015}{ New Journal of Physics \textbf{20}, 073015
  (2018).}

\bibitem{Braem-2018b}
B.~A. {Braem}, F.~M. {Pellegrino}, A.~{Principi}, M.~{R{\"o}{\"o}sli},
  S.~{Hennel}, J.~V. {Koski}, M.~{Berl}, W.~{Dietsche}, W.~{Wegscheider},
  M.~{Polini}, T.~{Ihn}, and K.~{Ensslin}, \emph{Scanning Gate Microscopy in a
  Viscous Electron Fluid}, 
  \href{https://journals.aps.org/prb/abstract/10.1103/PhysRevB.98.241304}
  {Phys. Rev. B \textbf{98}, 241304(R) (2018).}

\bibitem{Brun-thesis}
B.~Brun, \emph{Electron interactions in mesoscopic physics : Scanning Gate
  Microscopy and interferometry at a quantum point contact}, 
  \href{https://tel.archives-ouvertes.fr/tel-01137642}{Ph.{D}. thesis,
  {Universit{\'e} de Grenoble} (2014).}

\bibitem{Yacoby-1995}
A.~Yacoby, M.~Heiblum, D.~Mahalu, and H.~Shtrikman, \emph{Coherence and Phase
  Sensitive Measurements in a Quantum Dot}, 
  \href{http://link.aps.org/doi/10.1103/PhysRevLett.74.4047}{Phys. Rev. Lett. \textbf{74},
  4047--4050
  (1995).}

\bibitem{Zaffalon-2008}
M.~Zaffalon, A.~Bid, M.~Heiblum, D.~Mahalu, and V.~Umansky, \emph{Transmission
  Phase of a Singly Occupied Quantum Dot in the Kondo Regime},
  \href{http://link.aps.org/doi/10.1103/PhysRevLett.100.226601}{Phys. Rev. Lett.
  \textbf{100}, 226601
  (2008).}

\bibitem{Freyn-2008}
A.~Freyn, I.~Kleftogiannis, and J.-L. Pichard, \emph{Scanning Gate Microscopy
  of a Nanostructure Where Electrons Interact}, 
  \href{https://journals.aps.org/prl/abstract/10.1103/PhysRevLett.100.226802}
  {Phys. Rev. Lett. \textbf{100}, 226802 (2008).}

\bibitem{Soffing-2009}
S.~A. S\"offing, M.~Bortz, I.~Schneider, A.~Struck, M.~Fleischhauer, and
  S.~Eggert, \emph{Wigner crystal versus Friedel oscillations in the
  one-dimensional Hubbard model}, 
  \href{https://journals.aps.org/prb/abstract/10.1103/PhysRevB.79.195114}
  {Phys. Rev. B \textbf{79}, 195114 (2009).}

\bibitem{Puller-2004}
V.~I. Puller, L.~G. Mourokh, A.~Shailos, and J.~P. Bird, \emph{Detection of
  Local-Moment Formation Using the Resonant Interaction between Coupled Quantum
  Wires}, \href{https://link.aps.org/doi/10.1103/PhysRevLett.92.096802}{ 
  Phys. Rev. Lett. \textbf{92}, 096802 (2004).}

\bibitem{Steinberg-2006}
H.~Steinberg, O.~M. Auslaender, A.~Yacoby, J.~Qian, G.~A. Fiete,
  Y.~Tserkovnyak, B.~I. Halperin, K.~W. Baldwin, L.~N. Pfeiffer, and K.~W.
  West, \emph{Localization transition in a ballistic quantum wire},
  \href{https://link.aps.org/doi/10.1103/PhysRevB.73.113307}{Phys. Rev.
  B \textbf{73}, 113307
  (2006).}

\bibitem{Yoon-2007}
Y.~Yoon, L.~Mourokh, T.~Morimoto, N.~Aoki, Y.~Ochiai, J.~L. Reno, and J.~P.
  Bird, \emph{Probing the Microscopic Structure of Bound States in Quantum
  Point Contacts}, \href{https://link.aps.org/doi/10.1103/PhysRevLett.99.136805}{
  Phys. Rev. Lett. \textbf{99}, 136805
  (2007).}

\bibitem{Wu-2012}
P.~M. Wu, P.~Li, H.~Zhang, and A.~M. Chang, \emph{Evidence for the formation of
  quasibound states in an asymmetrical quantum point contact},
   Phys. Rev. B\href{https://link.aps.org/doi/10.1103/PhysRevB.85.085305}{
  \textbf{85}, 085305 (2012).}

\bibitem{Ho-2018}
S.-C. Ho, H.-J. Chang, C.-H. Chang, S.-T. Lo, G.~Creeth, S.~Kumar, I.~Farrer,
  D.~Ritchie, J.~Griffiths, G.~Jones, M.~Pepper, and T.-M. Chen, \emph{Imaging
  the Zigzag Wigner Crystal in Confinement-Tunable Quantum Wires}, 
  \href{https://link.aps.org/doi/10.1103/PhysRevLett.121.106801}{Phys. Rev.
  Lett. \textbf{121}, 106801
  (2018).}

\bibitem{Sushkov-2001}
O.~P. Sushkov, \emph{Conductance anomalies in a one-dimensional quantum
  contact},   \href{https://journals.aps.org/prb/abstract/10.1103/PhysRevB.64.155319}
  {Phys. Rev. B \textbf{64}, 155319 (2001).}

\bibitem{Sushkov-2003}
O.~P. Sushkov, \emph{Restricted and unrestricted Hartree-Fock calculations of
  conductance for a quantum point contact}, 
  \href{https://journals.aps.org/prb/abstract/10.1103/PhysRevB.67.195318}
  {Phys. Rev. B \textbf{67}, 195318 (2003).}

\bibitem{Shulenburger-2008}
L.~Shulenburger, M.~Casula, G.~Senatore, and R.~M. Martin, \emph{Correlation
  effects in quasi-one-dimensional quantum wires}, 
  \href{https://journals.aps.org/prb/abstract/10.1103/PhysRevB.78.165303}
  {Phys. Rev. B \textbf{78}, 165303 (2008).}

\bibitem{Guclu-2009}
A.~D. G\"u\ifmmode~\mbox{\c{c}}\else \c{c}\fi{}l\"u, C.~J. Umrigar, H.~Jiang,
  and H.~U. Baranger, \emph{Localization in an inhomogeneous quantum wire},
  \href{https://journals.aps.org/prb/abstract/10.1103/PhysRevB.80.201302}
  {Phys. Rev. B \textbf{80}, 201302 (2009).}

\bibitem{Berggren-2008}
K.-F. Berggren and I.~I. Yakimenko, \emph{Nature of electron states and
  symmetry breaking in quantum point contacts according to the local spin
  density approximation}, \href{http://stacks.iop.org/0953-8984/20/i=16/a=164203}{
  Journal of Physics: Condensed Matter \textbf{20}, 164203, (2008).}

\bibitem{Yakimenko-2013}
I.~I. Yakimenko, V.~S. Tsykunov, and K.-F. Berggren, \emph{Bound states,
  electron localization and spin correlations in low-dimensional GaAs/AlGaAs
  quantum constrictions}, \href{http://stacks.iop.org/0953-8984/25/i=7/a=072201}{Journal of Physics: Condensed Matter \textbf{25},
  072201
  (2013).}

\bibitem{Yoon-2009}
Y.~Yoon, M.-G. Kang, T.~Morimoto, L.~Mourokh, N.~Aoki, J.~L. Reno, J.~P. Bird,
  and Y.~Ochiai, \emph{Detector backaction on the self-consistent bound state
  in quantum point contacts}, \href{https://link.aps.org/doi/10.1103/PhysRevB.79.121304}{Phys. Rev. B \textbf{79}, 121304
  (2009).}

\bibitem{Ren-2010}
Y.~Ren, W.~W. Yu, S.~M. Frolov, J.~A. Folk, and W.~Wegscheider, \emph{Zero-bias
  anomaly of quantum point contacts in the low-conductance limit}, 
  \href{http://link.aps.org/doi/10.1103/PhysRevB.82.045313}{Phys. Rev. B \textbf{82}, 045313 (2010).}

\bibitem{Wigner-1934}
E.~Wigner, \emph{On the Interaction of Electrons in Metals}, Phys. Rev.
  \textbf{46}, 1002--1011 (1934).

\bibitem{Schimmel-2017}
D.~H. Schimmel, B.~Bruognolo, and J.~von Delft, \emph{Spin Fluctuations in the
  0.7 Anomaly in Quantum Point Contacts}, \href{https://link.aps.org/doi/10.1103/PhysRevLett.119.196401}{
  Phys. Rev. Lett. \textbf{119}, 196401 (2017).}

\bibitem{Kawamura-2015}
M.~Kawamura, K.~Ono, P.~Stano, K.~Kono, and T.~Aono, \emph{Electronic
  Magnetization of a Quantum Point Contact Measured by Nuclear Magnetic
  Resonance}, \href{https://link.aps.org/doi/10.1103/PhysRevLett.115.036601}
  {Phys. Rev. Lett. \textbf{115}, 036601 (2015).}

\bibitem{Note1}
\emph{These expressions are valid in the linear response regime. Though the
  experimental conditions do not satisfy the condition $\Delta T \ll T$, this
  linear model can partly explain the observations, which justifies \textit{a
  posteriori} this approximation.}

\bibitem{Cronenwett-thesis}
S.~M. Cronenwett, \emph{Coherence, charging and spin effects in quantum dots
  and quantum point contacts},\href{https://apps.dtic.mil/dtic/tr/fulltext/u2/1005432.pdf}
  {Ph.{D}. thesis, Harvard University, 2001.}

\bibitem{Iqbal-thesis}
M.~J. Iqbal, \emph{Electron many-body effects in quantum point contacts},
  \href{http://irs.ub.rug.nl/ppn/344115682}{Ph.{D}. thesis, Groningen University,
  2014}

\bibitem{Karavolas-1991}
V.~C. Karavolas and P.~N. Butcher, \emph{Diffusion thermopower of a 2DEG},
  \href{http://stacks.iop.org/0953-8984/3/i=15/a=016}{Journal of Physics: Condensed Matter \textbf{3}, 2597
  (1991).}

\bibitem{Fletcher-1995}
R.~Fletcher, P.~T. Coleridge, and Y.~Feng, \emph{Oscillations in the diffusion
  thermopower of a two-dimensional electron gas},\href{http://link.aps.org/doi/10.1103/PhysRevB.52.2823}{ Phys. Rev. B \textbf{52},
  2823--2830
  (1995).}

\bibitem{Schmidt-2012}
M.~Schmidt, G.~Schneider, C.~Heyn, A.~Stemmann, and W.~Hansen,
  \emph{Thermopower of a 2D Electron Gas in Suspended AlGaAs/GaAs
  Heterostructures}, \href{http://dx.doi.org/10.1007/s11664-011-1826-3}{Journal of Electronic Materials \textbf{41}, 1286-1289
  (2012).}

\bibitem{Tanatar-1989}
B.~Tanatar and D.~M. Ceperley, \emph{Ground state of the two-dimensional
  electron gas},\href{https://link.aps.org/doi/10.1103/PhysRevB.39.5005}{ Phys. Rev. B \textbf{39}, 5005--5016
  (1989).}

\end{thebibliography}


\newpage

\onecolumngrid

\includegraphics[page=1,scale=0.85]{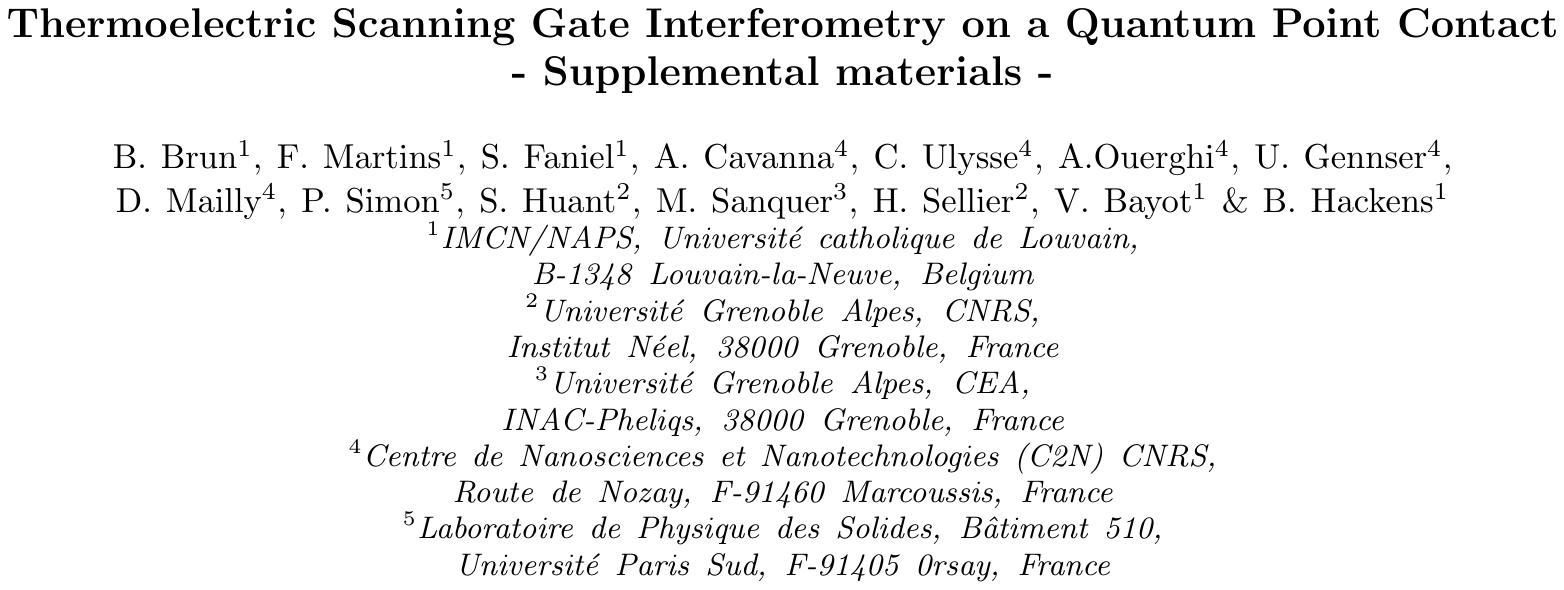} 
\includegraphics[page=2,scale=0.85]{Brun_supplemental.pdf} 
\includegraphics[page=3,scale=0.85]{Brun_supplemental.pdf} 
\includegraphics[page=4,scale=0.85]{Brun_supplemental.pdf} 
\includegraphics[page=5,scale=0.85]{Brun_supplemental.pdf} 
\includegraphics[page=6,scale=0.85]{Brun_supplemental.pdf} 
\includegraphics[page=7,scale=0.85]{Brun_supplemental.pdf} 
\includegraphics[page=8,scale=0.85]{Brun_supplemental.pdf} 

%

\end{document}